# Large spin accumulation voltages in epitaxial $Mn_5Ge_3$ contacts on Ge without oxide tunnel barrier

Aurélie Spiesser[*], Hidekazu Saito, Ron Jansen, Shinji Yuasa, and Koji Ando

*Spintronics Research Center, National Institute of Advanced Industrial Science and Technology (AIST), Tsukuba, Ibaraki 305-8568, Japan.*

Spin injection in high-quality epitaxial $Mn_5Ge_3$ Schottky contacts on n-type Ge has been investigated using a three-terminal Hanle effect measurement. Clear Hanle and inverted Hanle signals with features characteristic of spin accumulation and spin precession are observed up to 200 K. Strikingly, the observed spin voltage is several orders of magnitude larger than predicted by the theory of spin injection and diffusive spin transport. Since the devices have no oxide tunnel barrier, the discrepancy between theory and experiments cannot be explained by the often-invoked spin accumulation in localized states associated with the oxide or oxide/semiconductor interface. The observed spin voltages therefore must originate from the Ge itself, either from spins in the Ge bulk bands or its depletion region.

---

[*]Electronic mail: aurelie.spiesser@aist.go.jp

I. INTRODUCTION

The considerable progress in the electrical injection and detection of spin-polarized carriers in group-IV semiconductors has strengthened the field of semiconductor spintronics over the past few years.[1-17] To date, the most common technique to achieve efficient spin injection into a semiconductor (SC) such as silicon or germanium is to use a magnetic tunnel contact consisting of a conventional 3d-ferromagnet (FM) and an $Al_2O_3$, MgO or $SiO_2$ oxide tunnel barrier[1,3,5-7,9-14,16,17]. Subsequently, the presence of spin accumulation in the SC can be detected by electrical means using either a local three-terminal (3T) geometry[1,18] or a non-local (NL) four-terminal geometry[19,20]. The NL scheme uses two distinct ferromagnetic contacts for spin injection and detection, enabling the complete separation of the charge current path from the spin current path, provided that the devices are properly designed[20] such that there are no artifacts due to magnetic field-dependent offsets in the detector voltage. The three-terminal configuration consists of one single FM contact that acts as the spin injector as well as the spin detector. A first consequence is that the 3-terminal geometry is more sensitive to spurious signals such as Hall or anisotropic magneto-resistance (AMR) effects because of the non-zero charge current at the detector junction. However, these effects can be ruled out by a proper control experiment.[1,21] Secondly, theory work[22] shows that the spin voltage signal in the local geometry can have contributions not only from the spin accumulation induced in the semiconductor itself, but also from spin accumulation induced in localized states in the tunnel contact via two-step tunneling.[22]

As recently reviewed,[9,12] experimental data obtained by many research groups using the three-terminal configuration deviates significantly from what is expected on the basis of the theory of spin injection, accumulation and diffusion.[23] In particular, detected spin voltages many orders of magnitude larger than predicted by theory were reported for FM/oxide magnetic tunnel contacts on GaAs,[22] Si,[7,11,12,14,16] and Ge.[5,10,13,16] Furthermore, the scaling of the spin signal with the thickness of the tunnel barrier also disagrees with the theoretically expected trend.[16] Nonetheless, the spin signals, which are obtained from Hanle measurements, show all the characteristic features of precession of an induced non-equilibrium spin population. This indicates that interesting and yet to be uncovered physics is at play.

Although often overlooked, the magnitude of the spin signals detected in NL devices with a group-IV semiconductor channel also show deviations from the theoretically expected values. In Si- and Ge-based devices with Fe/MgO tunnel contacts, the experimentally observed NL spin signal is about two orders of magnitude smaller than expected, and in agreement with theory only if the tunnel spin polarization of the Fe/MgO contacts is taken to be a few percent,[3] or even less than one percent,[6] whereas a value of 50% is more reasonable. Interestingly, the same devices exhibit spin signals significantly larger than expected when the two-terminal magnetoresistance between the two ferromagnetic contacts is measured.[17] All this indicates that there is still a lack of quantitative understanding of the spin transport.

Returning to the 3T geometry, to explain the large spin voltage, the first and most often-cited theory involves two-step tunneling through localized states near the oxide/SC interface.[22,24] If the coupling of those states to the bulk semiconductor bands is sufficiently weak, a large spin accumulation may be induced in the interface states with a correspondingly enhanced spin voltage signal in a 3T Hanle measurement. While recent results indicate that this model does not reproduce the observed scaling with tunnel barrier thickness,[16] the debate is ongoing and new proposals have appeared.[14,25-27] Specifically, it was considered that the spin signal may originate from the oxide tunnel barrier itself,[14,22,25-27] possibly from localized states in the oxide close to the interface with the ferromagnet or from localized states deep within the tunnel oxide. More recently, the original theory[22] of two-step tunneling via localized states has been extended[26,27] by including the on-site Coulomb repulsion and considering the regime where the thermal energy is smaller than the energy range $eV$ of the tunnel electrons (set by the bias voltage $V$, with $e$ the electron charge), in which case back-flow into the source electrode is blocked and spins accumulate in the localized states unless forward tunneling into the other electrode is enabled by spin precession in an external magnetic field. We stress that all these explanations[14,22,25-27] involve localized states, either in the tunnel oxide or at the oxide/semiconductor interface, and also the existence of a non-equilibrium spin population, spin precession and the Hanle effect (even if a different name, i.e., resonant tunneling magnetoresistance, is used[26,27]).

Understanding the novel underlying physics of spin transport in 3T devices is an important goal, and it seems clear that the answer cannot be found in NL devices where the large signal enhancement is generally not observed. Considering this, we focus on 3T devices, and investigate another approach to create a spin accumulation in a SC by tunneling, namely via a FM/SC Schottky tunnel contact. This has been done previously using specific FM alloys such as $Fe_3Si$[2,15,28] or CoFe.[4,8] on Si and Ge with (111) orientation. Even though the growth of such alloys is usually hampered by the necessity to precisely control the compound stoichiometry,[29-31] a direct Schottky contact is an interesting system for 3T spin transport because there is no oxide tunnel barrier. This eliminates all possible enhancements of the spin voltage by localized states in the tunnel oxide or at the oxide/SC interface. Another important aspect is that in this work we use a native Schottky contact, i.e., we use a semiconductor with a homogeneous doping density. In contrast, most of the previous work on ferromagnet/semiconductor Schottky contacts has been carried out using a moderately doped semiconductor having a more heavily doped surface region[18,32-34] or a δ-doping layer near the surface.[2,4,8,15,28] The resulting non-trivial conduction band profile, that often includes a subsurface potential well, complicates the spin transport at the interface and therefore the interpretation of the results.[34-36]

Thus, we report here on spin injection in high-quality epitaxial $Mn_5Ge_3$ Schottky contacts on n-type Ge using a three-terminal Hanle effect measurement. The ferromagnetic $Mn_5Ge_3$ compound was grown epitaxially on Ge(111) using a simple and rapid growth technique and exhibited an abrupt interface with the Ge without any intermixing.[37,38] Clear Hanle and inverted Hanle signals with features characteristic of spin accumulation and spin precession are observed up to 200 K. The observed spin voltage is several orders of magnitude larger than predicted by the diffusive theory of spin injection and transport. Contrary to recent suggestions,[14,22,25-27] the discrepancy between theory and experiments cannot be explained by spin accumulation in localized states associated with the oxide or oxide/semiconductor interface since our devices do not have an oxide tunnel barrier. The observed spin voltages therefore must originate from the Ge itself, either from spins in the Ge bulk bands or its depletion region.

II. FABRICATION OF THE SCHOTTKY DEVICES

The epitaxial $Mn_5Ge_3$ Schottky contact was fabricated on an As-doped Ge(111) substrate with a carrier concentration of $1.1 \times 10^{18}$ cm$^{-3}$ and a resistivity of 6 mΩcm at 300 K. The resistance ($R$) of the substrate showed a very weak temperature dependence [$R$(10 K)/$R$(300 K) = 1.1], indicating a degenerate character. Prior to the deposition, the Ge substrate was first etched in dilute HF and then annealed at 700 °C for 10 min in an ultrahigh vacuum chamber to remove the native oxide. After this step, the Ge exhibited a well-ordered reconstruction pattern corresponding to a clean Ge surface.[38] Subsequently, we carried out the solid phase epitaxy method[37,38] by first depositing a 8 nm-thick Mn layer at room temperature (RT) using a Knudsen cell, and then annealing the sample in-situ at 450 °C to form the stable $Mn_5Ge_3$ compound. After this step, the reflection high-energy electron diffraction (RHEED) images displayed new streaky patterns, as shown in Figs. 1(a) and (b). These patterns correspond to the formation of a two dimensional epitaxial $Mn_5Ge_3$ film.[37] Finally, a 20 nm-thick Au cap layer was deposited at RT to prevent oxidation of the $Mn_5Ge_3$ film. The high-resolution cross-sectional transmission electron microscopy (TEM) image presented in Fig. 1(c) confirmed that an epitaxial $Mn_5Ge_3$ film with a high crystalline quality was obtained. The interface with Ge is free from any intermixing layers or clusters and is relatively abrupt. Nevertheless, in the low magnification TEM image (Fig. 1(d)), a non-negligible roughness is visible (the peak to valley roughness of the top surface exceeds 1 nm). Also, the contrast oscillations at the interface with Ge indicate a significant strain in the grown layer, which is due to a rather large lattice mismatch between Ge and $Mn_5Ge_3$ (about 3.7%).[37,38] The thickness of the $Mn_5Ge_3$ layer estimated from the TEM image was about 13 nm. The magnetic properties of the $Mn_5Ge_3$ film were measured by a superconducting quantum interference device (SQUID) magnetometer. A Curie temperature ($T_C$) of about 300 K was determined, which is consistent with previous reports.[37,38] For the transport measurements, standard micro-fabrication techniques (e.g., photolithography, Ar-ion milling, and $SiO_2$ sputtering) were carried out to fabricate the Schottky junctions with an active area of 100 × 200 $\mu m^2$. All the measurements were performed on the same device except for the spin signal versus bias voltage data presented in Fig. 6, which was obtained on a second device with nominally the same *I-V* characteristics and spin signals. To keep the measurement time within reasonable bounds, Hanle signals were measured at a limited number of different temperatures.

In previous 3T Hanle measurements using FM/oxide tunnel and Schottky contacts,[1,4-16,22,25,27] it is generally found that the Hanle signal $\Delta V$ is typically between 0.01 and 1% of the applied voltage *V*, so that the magnitude of the spin signal (*ΔV* divided by the current density) depends on the tunnel resistance. Although this scaling is not predicted by theory, there will be a particular value of the resistance for which the spin signal would, by coincidence, match the theoretically predicted value, which is determined by the parameters of the semiconductor (resistivity and spin-diffusion length) and the junction area. The Schottky devices studied here have a tunnel resistance that is orders of magnitude larger than the value for which such a coincidental agreement would occur, allowing a meaningful test of the agreement between theory and experiment.

III. RESULTS AND DISCUSSION

A. Schottky contact characteristics:

The current-voltage (*I-V*) characteristics of the junction measured at various temperatures are displayed in Fig. 2. The measurement was carried out between the junction and a Cr/Au contact with a large area (few mm$^2$) by a conventional two-probe method. All the *I-V* curves show a strong rectification, at all temperatures. This typical Schottky diode behavior indicates that the transport mechanism is not pure tunneling but thermally-assisted, which is in agreement with a previous study.[39] This feature is generally observed in direct metal contacts on n-type Ge because the Fermi level of the metal is strongly pinned near the top of the valence band of the Ge.[40,41] Although the mechanism underlying the Fermi level pinning at the metal/Ge interface is still unclear, one of the possible explanations is based on metal-induced gap states[42] that have energies within the band gap of the SC and penetrate from the metal into the SC.

B. Three-terminal Hanle effect measurements

To probe the presence of spin accumulation, we performed Hanle effect measurements in a 3T configuration.[1,18] As mentioned earlier, in this geometry the FM Schottky contact has the function of the spin injector and detector whereas two other contacts are used as the reference electrodes to apply a constant current and measure the voltage. The spin accumulation is thus detected in the SC just underneath the FM injector contact. For a normal Hanle measurement, a magnetic field ($B_\perp$) perpendicular to the film plane and thus transverse to the spins reduces the spin accumulation due to spin precession, inducing a change in the voltage across the junction, at constant current. However, the presence of local stray fields due to the roughness of the FM can induce a partial depolarization of the spins at zero magnetic field. This effect can be suppressed by applying an in-plane magnetic field ($B_{//}$), known as inverted Hanle effect.[43] Fig. 3(a) presents the change in the voltage across the $Mn_5Ge_3$/n-Ge Schottky contact measured at 15 K under forward bias and at a constant current of -40 µA. This corresponds to electron extraction from the Ge into the FM. For low transverse $B_\perp$ we observe a decrease of the voltage with increasing external magnetic field with a Lorentzian line shape. This can be ascribed to the Hanle precession of the electron spins that leads to a progressive suppression of the spin accumulation. On the other hand, when the external magnetic field is applied in the film plane ($B_{//}$), an increase of the voltage is observed, which corresponds to the inverted Hanle effect.[43] This is due to the recovery of the spin accumulation and indicates the presence of local magnetic fields in the junction. The total spin signal, which is defined as the sum of the voltage change in the Lorentzian part of the Hanle and the inverted Hanle curve is about 230 µV. No spin signal could be observed under reverse bias (electron injection condition) because the strong diode-like behavior limits the current across the Schottky junction and thereby the signal-to-noise ratio.

C. Signature of the Mn$_5$Ge$_3$ ferromagnet

The evolution of the Hanle signal (red curve) at higher external transverse field reflects the magnetic properties of the Mn$_5$Ge$_3$ film. As expected for a 13 nm-thick Mn$_5$Ge$_3$ film,[38,44] the magnetization at 15 K shows a hard axis perpendicular to the film plane (Fig. 3(b)). The magnetization has reached the fully out-of-plane orientation at a field of about $B_{sat}$ = 0.65 T. This behavior is in line with the Hanle curve showing first a reduction of the spin signal at low magnetic field due to spin precession, and then an upturn in the signal due to the rotation of the magnetization of the FM towards the out-of-plane direction. This reduces the spin precession. Above a magnetic field of about 0.65 T, the spin signal saturates because the magnetization, and therefore the spin of the injected electrons, is aligned with the external field. Notice that at high magnetic field, the Hanle and inverted Hanle signals saturate at different values, revealing an anisotropy between in-plane and out-of-plane magnetization. This has also been observed in FM/oxide/SC systems and is attributed to tunneling anisotropic magnetoresistance (TAMR) and/or tunneling anisotropic spin polarization.[45,46]

A new feature appears in the inverted Hanle curve in that it exhibits hysteresis. To better appreciate this, we present an enlargement of Fig. 3(a) at low magnetic field in Fig. 4(a). The presence of a hysteresis in the inverted Hanle curves can be clearly seen around $B$ = ± 50 mT. As previously explained,[43] the inverted Hanle effect is the consequence of the local magnetostatic fields that originate from the roughness of the FM. A sizeable roughness was indeed revealed by the TEM observations (Fig. 1(d)). However, the observation of a strong hysteresis suggests that another source of local magnetostatic fields exists. The magnetic properties of the Mn$_5$Ge$_3$ film for the in-plane applied field at 15 K are displayed in Fig. 4(b). The curve exhibits a square-like shape, confirming that the easy axis of the magnetization lies in the plane of the film.[37,47] More importantly, as already reported for such thin Mn$_5$Ge$_3$ nanostructures,[44,48] the shape of the curve is consistent with the process of magnetization reversal by domain nucleation. This induces the formation of domain walls that are Bloch-type, i.e., with a magnetization perpendicular to the film plane, which can causes additional stray field around the coercive force of ± 50 mT. Therefore, we attribute the presence of hysteresis in the inverted Hanle curve to the variation of the local stray field during the magnetization reversal of the Mn$_5$Ge$_3$ film. In a previous optical study of spin precession in Ni/GaAs structures, the effect of stray fields created by the domain structure of the ferromagnet was also reported.[49] At large $B_{//}$ we observed that the voltage across the junction saturates because spin precession in the local magnetostatic fields is suppressed. Note that for a process of magnetization reversal by rotation, the magnetization and thus the injected spins would make an angle with the applied field. This would result in spin precession and would also produce hysteresis.

D. Evolution of the spin signal with the temperature

Clear Hanle and inverted Hanle effects could be still observed up to 200 K under forward bias, as shown in the Fig. 5(a). Closer to room temperature, no spin signal could be detected, which can be expected for a FM with a $T_C$ of about 300 K. Recently, an enhancement of the $T_C$ of Mn$_5$Ge$_3$ up to 430 K was demonstrated by using a carbon-doping technique while keeping a high crystalline quality of the structure and a high thermal stability.[50,51] Hence, this could lead to the creation of spin accumulation at RT in carbon-doped Mn$_5$Ge$_3$/n-Ge junctions.

Next, the junction-area-product [junction-$RA = (V / I) \times A$] and the spin resistance-area-product [spin-$RA = (\Delta V / I) \times A$] are plotted as a function of temperature ($T$) in Fig. 5(b). All the measurements were taken under forward bias at the same bias voltage. An exponential decay is observed with increasing temperature up to 200 K, with the decay rate of the spin-*RA* being slightly larger than that of the junction-*RA*. The spin signal is proportional to $P^2$, where $P$ is the tunneling spin polarization, and at temperatures well below $T_C$, $P$ is expected to decay with $T$ as $1 - \alpha T^{3/2}$ with an $\alpha$ that depends on the $T_C$ and on the junction interface properties.[52] For a tunnel interface, the parameter $\alpha$ is generally larger than the one that describes the decay of the bulk magnetization with temperature.[52]

From the normal Hanle curves, we can estimate the effective spin lifetime ($\tau_s$) by using the following expression: $\Delta V(B_\perp) = \Delta V(0) / [1 + (\omega_L \tau_s)^2]$, where $\omega_L$ is the Larmor frequency ($\omega_L = g\mu_B/\hbar$, where $g$ is the electron g-factor ($g = 1.6$ for Ge),[53] $\mu_B$ is the Bohr magneton and $\hbar$ is Planck's constant divided by $2\pi$). The fitting curve is represented by the solid line in Fig. 4(a) and Fig. 5(a) and gives a $\tau_s$ of 146 ps and 136 ps at 15 K and 200 K, respectively. These values are noticeably lower than the values (several ns) extracted from electron spin resonance on As-doped Ge with a similar doping density,[54] but we should recall that the presence of a sizeable inverted Hanle effect implies that additional magnetic fields are present, causing an artificial broadening of the Hanle curve and thereby an apparent reduction of the spin lifetime.[43]

E. Evolution of the spin signal with bias voltage

Fig. 6 presents the evolution of the absolute current $|I|$, the spin signal $\Delta V$, the junction-*RA* and the spin-*RA* as a function of the forward bias voltage $V$ ($V < 0$) at 15 K. The salient features are as follows. Firstly, the current $|I|$ rises exponentially with increasing $|V|$ (Fig. 6(a)), which is typical for a Schottky diode under forward bias. As a consequence, the junction-*RA* decreases exponentially with bias, as shown in Fig. 6(c). Secondly, the spin signal $\Delta V$ does not increase as a function of bias voltage, but instead exhibits a weak decay (Fig. 6 (b)). The total spin signal $\Delta V$, obtained from the sum of the Hanle and inverted Hanle signal amplitudes, has a value of about 250 μV at $V = -250$ mV and decreases quasi-linearly to a value of 180 μV at $V = -425$ mV. The signal corresponding to the Hanle contribution only (green diamonds in Fig. 6 (b)) also exhibits a quasi-linear decrease. Thirdly, the spin-$RA$ product decays exponentially as a function of bias voltage (Fig. 6 (c)) and the decay is slightly faster than that of the junction-$RA$ product. Although some decay of the spin signal with bias is expected due to the known reduction of the tunnel spin polarization $P$ at higher energy, the expected decay is rather weak.[55] In fact, the close resemblance of the behavior of the spin-$RA$ and the junction-$RA$ product suggests that the spin voltage $\Delta V$ is not proportional to the current but to the applied voltage (a constant $\Delta V/V$ is equivalent to a constant ratio of the spin-$RA$ and the junction-$RA$ product). A similar scaling was recently reported for 3T measurements on FM/oxide tunnel contacts on Si and Ge where the thicknesses of the oxide was varied.[16] We note that because our device has a strongly non-linear $I$-$V$ curve, a study of the spin signal versus bias voltage can be used to distinguish between scaling of the $\Delta V$ with the current or with the voltage (this is not possible in a device with a linear $I$-$V$ curve).

F. Comparison with theory

We now compare the experimental value of the spin-$RA$ to the value predicted from the standard theory for spin injection in a nonmagnetic material.[23] The theoretical spin-$RA$ is equal to $P^2\rho_{Ge}l_{sd}$ where $P$ is the tunnel spin polarization of $Mn_5Ge_3$, $\rho_{Ge}$ the resistivity of the Ge substrate and $l_{sd}$ the spin-diffusion length. Taking $P \sim 0.5$ and $\rho_{Ge} = 0.006$ Ωcm at 15 K, and assuming $l_{sd} = 1$ μm, we obtain a theoretical spin-$RA$ of 15 Ωμm². The measured spin-$RA$ is 100 kΩμm² at 15 K (Fig. 5(b)) and even larger at lower bias (Fig. 6(c)), and thus about 4 orders of magnitude larger than the theoretical value. Moreover, the spin voltage is comparable to that previously observed for FM/oxide/Ge tunnel junctions.[5,10,13,16] We conclude that despite the absence of the oxide tunnel barrier, the spin voltage deviates significantly from the theoretically expected value by many orders of magnitude. Moreover, we observed a scaling of the spin-$RA$ product with the junction-$RA$ product when the temperature (Fig. 5(b)) or the bias voltage (Fig. 6(c)) is varied. This scaling is, as previously pointed out,[16] also not consistent with the predictions of the theory. Next we discuss the origin of the discrepancy.

G. Discussion of the origin of the large spin voltage

   1) States related to the oxide

It has frequently been argued that the enhanced Hanle spin signals originate from localized states at the oxide/SC interface or originate from the oxide barrier itself,[14,22,25-27] that this can explain all the spin signals observed on FM/oxide/SC devices obtained with the 3T geometry,[26,27] and that the spin signals do not arise from spins in the non-magnetic channel.[26,27] However, since the Schottky contacts studied here do not have an oxide tunnel barrier, the enhanced spin signal cannot be explained by any mechanism that involves the oxide and the localized states these may produce. We can thus exclude the explanations given by Tran *et al.*[22] (two-step tunneling via localized states at the oxide/semiconductor interface), Uemura *et al.*[14] (localized states in the oxide near the FM interface), Txoperena *et al.*[25] (states in the oxide barrier), as well as the "resonant tunnel magnetoresistance" proposed by Song *et al.*[26] (impurity defect states in the tunnel oxide) and later also invoked by Txoperena *et al.*.[27] Note that in the latter case metallic junctions were used with oxide tunnel barriers that were deliberately made to be oxygen-deficient to enable the impurity-assisted mechanism to appear, whereas the associated spin signals were absent in well-oxidized tunnel barriers created by plasma oxidation. Finally, we note that interface states, i.e., metal-induced gap states, are certainly expected to be present in FM/SC Schottky contacts. However, such states are directly coupled to the FM and hence these interface states cannot sustain a large spin accumulation as spins will be drained away by the FM that acts as an efficient spin sink.

2) Conduction band profile

As mentioned in the introduction, in ferromagnet/semiconductor Schottky contacts one commonly uses a semiconductor with a surface doping profile, either a graded profile with a more heavily doped surface region[18,32-34] or a δ-doping layer near the surface.[2,4,8,15,28] The non-homogeneous doping density results in a inhomogeneity of the spin-transport parameters of the semiconductor (spin-relaxation time, spin-diffusion length, etc.). More importantly, the conduction band profile is non-trivial and often includes a subsurface potential well. All this complicates the spin transport at the interface and prevents a quantitative comparison between experiment and theory.[34-36] Noteworthy is that the spin accumulation, when confined to the subsurface potential well, can be enhanced and with it the spin signals that are detected in a 3T measurement.

In this regard it is important to keep in mind that in the work presented here we did not use a doping profile, but instead we studied Schottky contacts on a semiconductor with a homogeneous doping density. The abovementioned complications are therefore absent in our devices and we can thus exclude any possible source of spin signal enhancement associated with a surface doping profile.

3) Effect of spin drift

In the theory work by Yu *et al.*[56] it was pointed out that spin injection from a FM into a SC can be affected by spin drift due to the electric field that is induced in the SC. Recently, the effect of spin drift was investigated in devices containing FM/MgO tunnel contacts on Si.[57] Whereas no spin drift effects were observed for 4-terminal non-local measurements, it was argued that spin drift effects are important for 3T measurements.[57] This raises the question whether this effect is relevant in our Ge-based Schottky devices. The low resistivity of the Ge substrate used, its large thickness (about 0.3 mm) and the small current density due to the large area of the contact limit the electric field in the Ge to several tens of mVcm$^{-1}$ for a bias voltage of -300 mV. At such small values of the electric field, spin drift is negligible.[56] Therefore, the effect of spin drift cannot explain the enhancement of the spin accumulation by several orders of magnitude.

4) Non-linearity of the *I-V* characteristics

It has recently been pointed out that the magnitude of the spin signals can be modified if the transport across the contact is non-linear.[58,59] It is relevant to discuss this possibility, since the Schottky contacts we studied here are highly rectifying and exhibit strongly non-linear *I-V* characteristics. Let us first comment on how the non-linearity affects the spin signal. In previous works,[58,59] it was argued that when transport is non-linear, the spin signal is enhanced by a factor $(dV/dI)/(V/I)$ representing the ratio of the differential resistance and the resistance of the contact. We have recently found that in general one cannot describe the effect of the non-linearity on the spin signal by this simple multiplication.[60] Explicit theoretical evaluation[60] of spin signals for thermally-activated transport across a Schottky barrier shows that the ratio of differential resistance and resistance does not appear in the expression for the spin signal, even though the transport is highly non-linear and rectifying (note that for a Schottky diode under forward bias, ($dV/dI$)/($V/I$) < 1, which according to previous works[58,59] would produce a reduction of the spin signal due to the non-linearity). Secondly, it was found[60] that a non-linearity, characterized by a non-zero $d^2I/dV^2$, changes the magnitude of the detected spin signal if the magnitude $\Delta\mu$ of the induced spin accumulation is sufficiently large compared to the characteristic energy scale $E_0$ that describes the degree of non-linearity. When the spin accumulation is, as usual, small, the transport parameters are essentially constant over the energy range of the spin accumulation. The spin accumulation is then not able to probe the non-linearity. The parameter $E_0$ can be obtained by fitting the forward bias part of the experimental I-V curves (Fig. 2) to the following expression: $I \propto \exp(q|V|/E_0)$. The result is $E_0 = 25$ meV, which indeed is much larger than the value of the spin accumulation. Moreover, even if the spin accumulation is large compared to $E_0$, it will in general not produce an enhancement of the spin signal but instead a reduction, because I-V curves are typically super-linear (with a conductance that increases with bias voltage). The spin-detection sensitivity is enhanced in special cases for which the conductance decreases with bias voltage. We conclude that the large spin signals observed in our Schottky device are not due to the non-linearity of the electrical transport.

5) Spins in the Ge

Let us first of all stress that notwithstanding the signal magnitude, the spin voltage obtained from the Hanle measurements shows all the characteristic features of precession of an induced non-equilibrium spin population: a signal decay with a Lorentzian shape for small magnetic fields perpendicular to the spins, a signal recovery for larger perpendicular fields due to rotation of the ferromagnetic injector into the direction of the applied field, and an inverted Hanle effect for applied magnetic fields parallel to the spin direction.[61] This also allows us to rule out TAMR, AMR and Hall voltages since these do not produce a signal with a Lorentzian line shape. And since the signal cannot originate from any oxide-related localized states, the large spin signal can only come from spins in the Ge itself.

Still, there are two possibilities. One is that the spin signal is due to a genuine spin accumulation in the Ge conduction band, and that this produces a signal much larger than predicted by theory for a not yet known reason. Alternatively, the signal may come from the depletion region in the Ge (that forms the tunnel barrier) if it contains defect states. It certainly contains shallow dopant states, but these are strongly coupled to the conduction band and hence cannot sustain an enhanced spin accumulation (for each spin, their occupation is described by the same electrochemical potential as the bulk conduction band states). Therefore, we consider that it is possible that some Mn atoms may have diffused into the Ge during the growth process. Since Mn can act as a double acceptor in Ge with energy levels from the valence band at + 0.16 eV and from the conduction band at - 0.32 eV (mid-gap), two-step tunneling through the Mn impurities in the depletion region of the Ge can occur. Although this cannot be confirmed or excluded with the data at hand, it would still imply that the large spin precession signal originates from the Ge. Further work is needed to clarify whether or not Mn defects in the depletion region play a role.

IV. CONCLUSIONS

To conclude, we have investigated electrical spin injection using an epitaxial $Mn_5Ge_3$ Schottky contact on n-type Ge(111), where the $Mn_5Ge_3$ film exhibits a high crystalline quality and a well-defined interface with the Ge. Hanle and inverted Hanle effects were clearly detected up to 200 K using the 3T geometry. Despite the absence of an oxide tunnel barrier, the observed spin signal is 4 orders of magnitude larger than the predicted value. By using a native Schottky contact without tunnel oxide, we can rule out any signal enhancement due to mechanisms that involve localized states in an oxide or at the oxide/SC interface. Since the Hanle signals have all the features characteristic of spin accumulation and spin precession, we conclude that the observed spin voltages must originate from the Ge itself, either from spins in the Ge bulk bands or its depletion region.

ACKNOWLEDGMENTS

This work was supported by the Japan Society for the Promotion of Science (JSPS) Postdoctoral Fellowship for Foreign Researchers (No. 2301816) and the Funding Program for Next Generation World-Leading Researchers (No. GR099).

REFERENCES

[1] S. P. Dash, S. Sharma, R. S. Patel, M. P. de Jong, and R. Jansen, Nature **462**, 491 (2009).

[2] Y. Ando, K. Hamaya, K. Kasahara, Y. Kishi, K. Ueda, K. Sawano, T. Sadoh, and M. Miyao, Appl. Phys. Lett. **94**, 182105 (2009).

[3] T. Suzuki, T. Sasaki, T. Oikawa, M. Shiraishi, Y. Suzuki, and K. Noguchi, Appl. Phys. Express **4**, 023003 (2011).

[4] Y. Ando, K. Kasahara, K. Yamane, Y. Baba, Y. Maeda, Y. Hoshi, K. Sawano, M. Miyao, and K. Hamaya, Appl. Phys. Lett. **99**, 012113 (2011).

[5] K.-R. Jeon, B.-C. Min, Y.-H. Jo, H.-S. Lee, I.-J. Shin, C.-Y. Park, S.-Y. Park, and S.-C. Shin, Phys. Rev. B **84**, 165315 (2011).

[6] Y. Zhou, W. Han, L.-T. Chang, F. Xiu, M. Wang, M. Oehme, I. A. Fischer, J. Schulze, R. K. Kawakami, and K. L. Wang, Phys. Rev. B **84**, 125323 (2011).

[7] C. H. Li, O. M. J. van't Erve, and B. T. Jonker, Nat. Commun. **2**, 245 (2011).

[8] Y. Ando, K. Kasahara, S. Yamada, Y. Maeda, K. Masaki, Y. Hoshi, K. Sawano, M. Miyao, and K. Hamaya, Phys. Rev. B **85**, 035320 (2012).

[9] R. Jansen, Nat. Mater. **11**, 400 (2012).

[10] S. Iba, H. Saito, A. Spiesser, S. Watanabe, R. Jansen, S. Yuasa, and K. Ando, Appl. Phys. Express **5**, 053004 (2012).

[11] M. Ishikawa, H. Sugiyama, T. Inokuchi, K. Hamaya, and Y. Saito, Appl. Phys. Lett. **100**, 252404 (2012).

[12] R. Jansen, S. P. Dash, S. Sharma, and B. C. Min, Semicond. Sci. Technol. **27**, 083001 (2012).

[13] A. Jain, C. Vergnaud, J. Peiro, J. C. Le Breton, E. Prestat, L. Louahadj, C. Portemont, C. Ducruet, V. Baltz, A. Marty, A. Barski, P. Bayle-Guillemaud, L. Vila, J.-P. Attane, E. Augendre, H. Jaffrès, J.-M. George, and M. Jamet, Appl. Phys. Lett. **101**, 022402 (2012).

[14] T. Uemura, K. Kondo, J. Fujisawa, K.-I. Matsuda, and M. Yamamoto, Appl. Phys. Lett. **101**, 132411 (2012).

[15] K. Hamaya, Y. Baba, G. Takemoto, K. Kasahara, S. Yamada, K. Sawano, and M. Miyao, J. Appl. Phys. **113**, 183713 (2013).

[16] S. Sharma, A. Spiesser, S. P. Dash, S. Iba, S. Watanabe, B. J. van Wees, H. Saito, S. Yuasa, and R. Jansen, Phys. Rev. B **89**, 075301 (2014).

[17] T. Sasaki, T. Suzuki, Y. Ando, H. Koike, T. Oikawa, Y. Suzuki, and M. Shiraishi, Appl. Phys. Lett. **104**, 052404 (2014).

[18] X. Lou, C. Adelmann, M. Furis, S. Crooker, C. Palmstrøm, and P. Crowell, Phys. Rev. Lett. **96**, 176603 (2006).

[19] M. Johnson and R. H. Silsbee, Phys. Rev. B **35**, 4959 (1987).

[20] F. J. Jedema, A. T. Filip, and B. J. van Wees, Nature **410**, 345 (2001).

[21] R. S. Patel, S. P. Dash, M. P. de Jong, and R. Jansen, J. Appl. Phys. **106**, 016107 (2009).

[22] M. Tran, H. Jaffrès, C. Deranlot, J. George, A. Fert, A. Miard, and A. Lemaître, Phys. Rev. Lett. **102**, 036601 (2009).

[23] A. Fert and H. Jaffrès, Phys. Rev. B **64**, 184420 (2001).

[24] R. Jansen, A. M. Deac, H. Saito, and S. Yuasa, Phys. Rev. B **85**, 134420 (2012).

[25] O. Txoperena, M. Gobbi, A. Bedoya-Pinto, F. Golmar, X. Sun, L. E. Hueso, and F. Casanova, Appl. Phys. Lett. **102**, 192406 (2013).

[26] Y. Song and H. Dery, Phys. Rev. Lett. **113**, 047205 (2014).

[27] O. Txoperena, Y. Song, L. Qing, M. Gobbi, L. E. Hueso, H. Dery and F. Casanova, arXiv:1404.0633v1 (2014).

[28] Y. Ando, K. Kasahara, K. Yamane, K. Hamaya, K. Sawano, T. Kimura, and M. Miyao, Appl. Phys. Express **3**, 093001 (2010).

[29] T. Sadoh, M. Kumano, R. Kizuka, K. Ueda, a. Kenjo, and M. Miyao, Appl. Phys. Lett. **89**, 182511 (2006).

[30] K. Hamaya, K. Ueda, Y. Kishi, Y. Ando, T. Sadoh, and M. Miyao, Appl. Phys. Lett. **93**, 132117 (2008).

[31] Y. Maeda, K. Hamaya, S. Yamada, Y. Ando, K. Yamane, and M. Miyao, Appl. Phys. Lett. **97**, 192501 (2010).

[32] A. Fuhrer, S. F. Alvarado, G. Salis, and R. Allenspach, Appl. Phys. Lett. **98**, 202104 (2011).

[33] T. Uemura, T. Akiho, M. Harada, K. Matsuda, and M. Yamamoto, Appl. Phys. Lett. **99,** 082108 (2011).

[34] Q. O. Hu, E. S. Garlid, P. A. Crowell, and C. J. Palmstrøm, Phys. Rev. B **84**, 085306 (2011).

[35] H. Dery and L. J. Sham, Phys. Rev. Lett. **98**, 046602 (2007).

[36] Y. Song and H. Dery, Phys. Rev. B **81**, 045321 (2010).

[37] C. Zeng, S. C. Erwin, L. C. Feldman, a. P. Li, R. Jin, Y. Song, J. R. Thompson, and H. H. Weitering, Appl. Phys. Lett. **83**, 5002 (2003).

[38] S. Olive-Mendez, A. Spiesser, L. A. Michez, V. Le Thanh, A. Glachant, J. Derrien, and T. Devillers, Thin Solid Films **517**, 191 (2008).

[39] A. Sellai, A. Mesli, M. Petit, V. Le Thanh, D. Taylor, and M. Henini, Semicond. Sci. Technol. **27**, 035014 (2012).

[40] A. Dimoulas, P. Tsipas, A. Sotiropoulos, and E. K. Evangelou, Appl. Phys. Lett. **89**, 252110 (2006).

[41] T. Nishimura, K. Kita, and A. Toriumi, Appl. Phys. Lett. **91**, 123123 (2007).

[42] T. Nishimura, K. Kita, and A. Toriumi, Appl. Phys. Express **1**, 051406 (2008).

[43] S. P. Dash, S. Sharma, J. C. Le Breton, J. Peiro, H. Jaffrès, J.-M. George, A. Lemaître, and R. Jansen, Phys. Rev. B **84**, 054410 (2011).

[44] A. Spiesser, F. Virot, L. Michez, R. Hayn, S. Bertaina, L. Favre, M. Petit, and V. Le Thanh, Phys. Rev. B **86**, 035211 (2012).

[45] S. Sharma, S. P. Dash, H. Saito, S. Yuasa, B. J. van Wees, and R. Jansen, Phys. Rev. B **86**, 165308 (2012)

[46] S. Sharma, A. Spiesser, H. Saito, S. Yuasa, B. J. van Wees, and R. Jansen, Phys. Rev. B **87**, 085307 (2013).

[47] A. Spiesser, S. F. Olive-Mendez, M. Dau, L. A. Michez, A. Watanabe, V. Le Thanh, and A. Glachant, Thin Solid Films **518**, S113 (2010).

[48] J. Tang, C.-Y. Wang, W. Jiang, L.-T. Chang, Y. Fan, M. Chan, C. Wu, M.-H. Hung, P.-H. Liu, H.-J. Yang, H.-Y. Tuan, L.-J. Chen, and K. L. Wang, Nano Lett. **12**, 6372 (2012).

[49] R. I. Dzhioev, V. L. Korenev, and B. P. Zakharchenya, Solid State Phys. **37**, 1929 (1995).

[50] A. Spiesser, V. Le Thanh, S. Bertaina, and L. A. Michez, Appl. Phys. Lett. **99**, 121904 (2011).

[51] A. Spiesser, I. Slipukhina, M. Dau, E. Arras, V. Le Thanh, L. A. Michez, P. Pochet, H. Saito, S. Yuasa, M. Jamet, and J. Derrien, Phys. Rev. B **84**, 165203 (2011).

[52] C. H. Shang, J. Nowak, R. Jansen, and J. Moodera, Phys. Rev. B **58**, R2917 (1998).

[53] G. Feher, D. K. Wilson, and E. Gere, Phys. Rev. Lett. **3**, 25 (1959).

[54] E. M. Gershenzon, N. M. Pevin, and M. S. Fouelson, Phys. Status Solidi **49**, 287 (1972).

[55] S. Valenzuela, D. Monsma, C. Marcus, V. Narayanamurti, and M. Tinkham, Phys. Rev. Lett. **94**, 196601 (2005).

[56] Z. Yu and M. Flatté, Phys. Rev. B **66**, 235302 (2002).

[57] M. Kameno, Y. Ando, E. Shikoh, T. Shinjo, T. Sasaki, T. Oikawa, Y. Suzuki, T. Suzuki, and M. Shiraishi, Appl. Phys. Lett. **101**, 122413 (2012). Note that in this work, the authors only reported the decay of the effective spin-diffusion length for one bias polarity, but not the complementary increase of the spin-diffusion length that is expected for the opposite electric field polarity in the case of spin drift.

[58] Y. Pu, J. Beardsley, P. M. Odenthal, A. G. Swartz, R. K. Kawakami, P. C. Hammel, E. Johnston-Halperin, Jairo Sinova, and J. P. Pelz, Appl. Phys. Lett. **103**, 012402 (2013).

[59] J. Shiogai, M. Ciorga, M. Utz, D. Schuh, M. Kohda, D. Bougeard, T. Nojima, J. Nitta, and D. Weiss, Phys. Rev. B **89**, 081307 (2014).

[60] R. Jansen, A. Spiesser, H. Saito, S. Yuasa, arXiv:1410.3994 (2014).

[61] Note that for a recent work (Aoki *et al.*, Phys. Rev. B **86**, 081201 (2012)), it is not clear whether the magnetic field dependent voltage signals observed using the 3T geometry are due to a Hanle effect or due to an artifact unrelated to spin accumulation. The recovery of the

voltage signal due to rotation of the ferromagnet at higher perpendicular magnetic field was not investigated, the signal did not appear to scale with the tunnel current, and no control experiments were performed to rule out well-known artifacts due to AMR, TAMR or Hall voltages from stray fields.

**FIGURE CAPTIONS**

FIG. 1 RHEED patterns of the $Mn_5Ge_3$ layer along the (a) [1-10] and (b) [11-2] azimuths of the Ge substrate, respectively. (c) High-resolution and (d) low magnification cross-sectional TEM image of the Au/ $Mn_5Ge_3$/n-Ge structure.

FIG. 2 Current-voltage characteristics of the $Mn_5Ge_3$/n-Ge device measured at various temperatures. The bias voltage is defined as $V_{Ge} - V_{Mn5Ge3}$, where $V_{Ge}$ and $V_{Mn5Ge3}$ are the potentials of the Ge and $Mn_5Ge_3$ electrodes, respectively.

FIG. 3 (a) Hanle ($B_\perp$) and inverted Hanle ($B_{//}$) curves of the $Mn_5Ge_3$/n-Ge device measured at 15 K with $V \sim$ - 300 mV and $I$ = - 40 µA. This bias convention corresponds to electron extraction from the Ge into the FM (forward bias). Blue and green curves correspond to the field swept from positive to negative and from negative to positive values, respectively. (b) Magnetization curve measured at 15 K with the magnetic field applied in the film plane ($B_\perp$).

FIG. 4 (a) Enlargement of the Hanle ($B_\perp$) and inverted Hanle ($B_{//}$) curves of the $Mn_5Ge_3$/n-Ge device measured at 15 K with $V \sim$ - 300 mV and $I$ = - 40 µA under forward bias. Blue and green curves correspond to the field swept from positive to negative and from negative to positive values, respectively. (b) Magnetization curve measured at 15 K with the magnetic field applied in the film plane ($B_{//}$).

FIG. 5 (a) Hanle ($B_\perp$) and inverted Hanle ($B_{//}$) curves of the $Mn_5Ge_3$/n-Ge device at 200 K with $V$ ~ - 300 mV and $I$ = - 3.37 mA under forward bias. (b) Junction-$RA$ and spin-$RA$ as a function of the temperature measured at a bias voltage of $V$ ~ - 300 mV.

FIG. 6 (a) Absolute current $|I|$, (b) spin voltage $\Delta V$ and (c) junction- and spin-$RA$ product of a $Mn_5Ge_3$/n-Ge device at 15 K as a function of the forward bias voltage $V$. In Fig. (b), the orange triangles and green diamonds correspond to $\Delta V_{Total}$ and $\Delta V_{Hanle}$, respectively, where $\Delta V_{Total}$ is obtained from the sum of the Hanle and inverted Hanle signal amplitudes. The data was obtained on a different (yet nominally identical) device than the one used for Figs. 2 – 5. In Fig. 6(c), the two black data points at $V$ = - 300 mV correspond to the $Mn_5Ge_3$/n-Ge device that was used for Figs. 2 – 5.

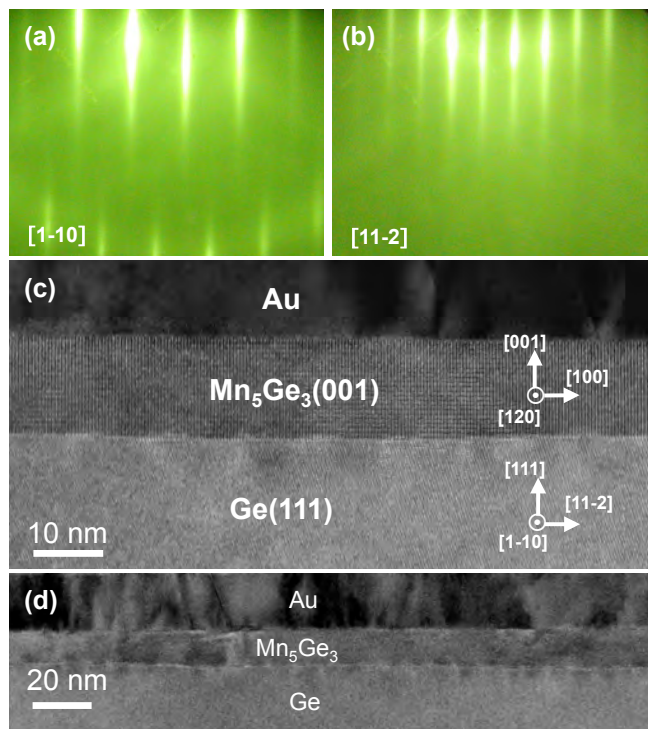

FIG. 1 (a), (b), (c) and (d)

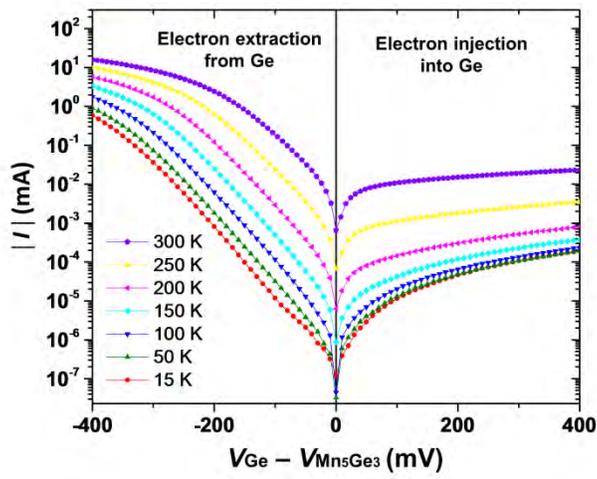

FIG. 2

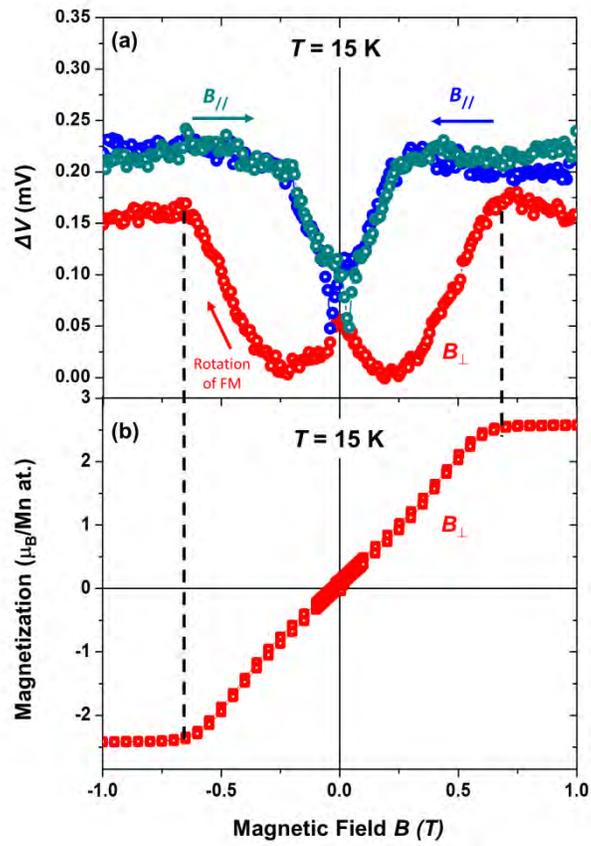

FIG. 3 (a) and (b)

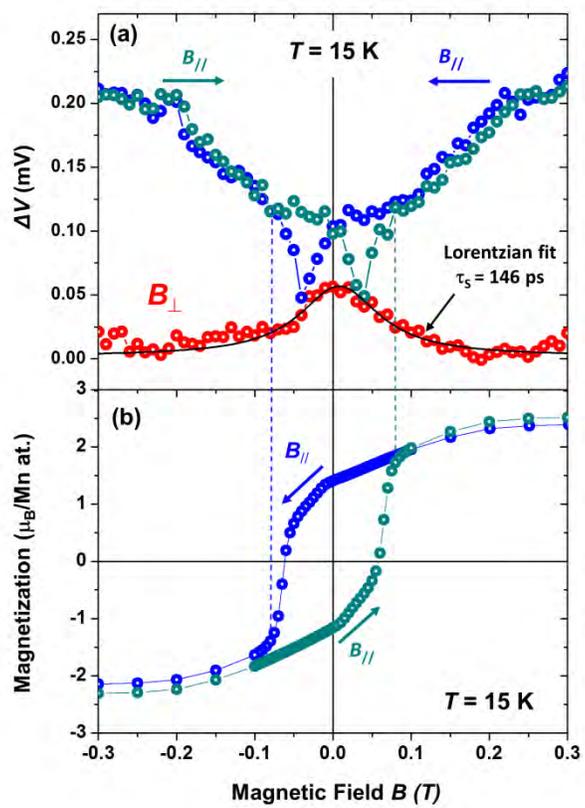

FIG. 4 (a) and (b)

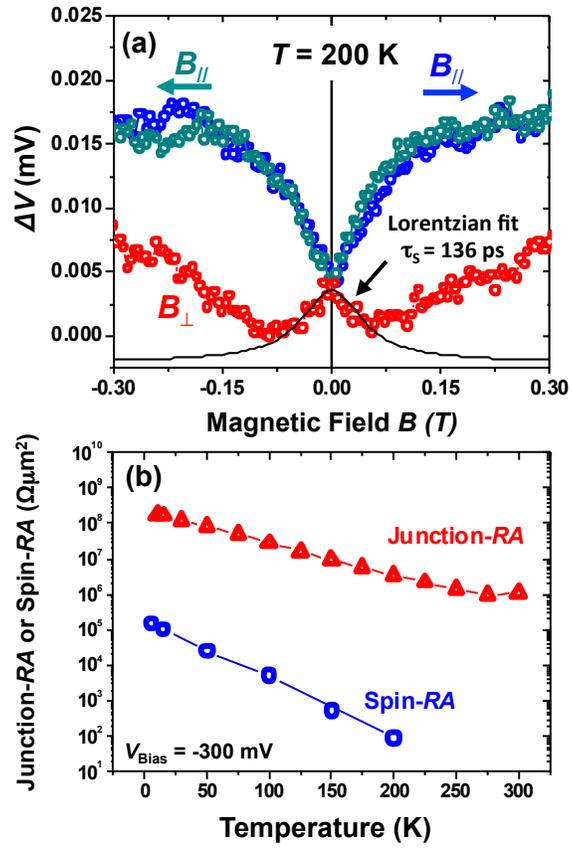

FIG. 5 (a) and (b)

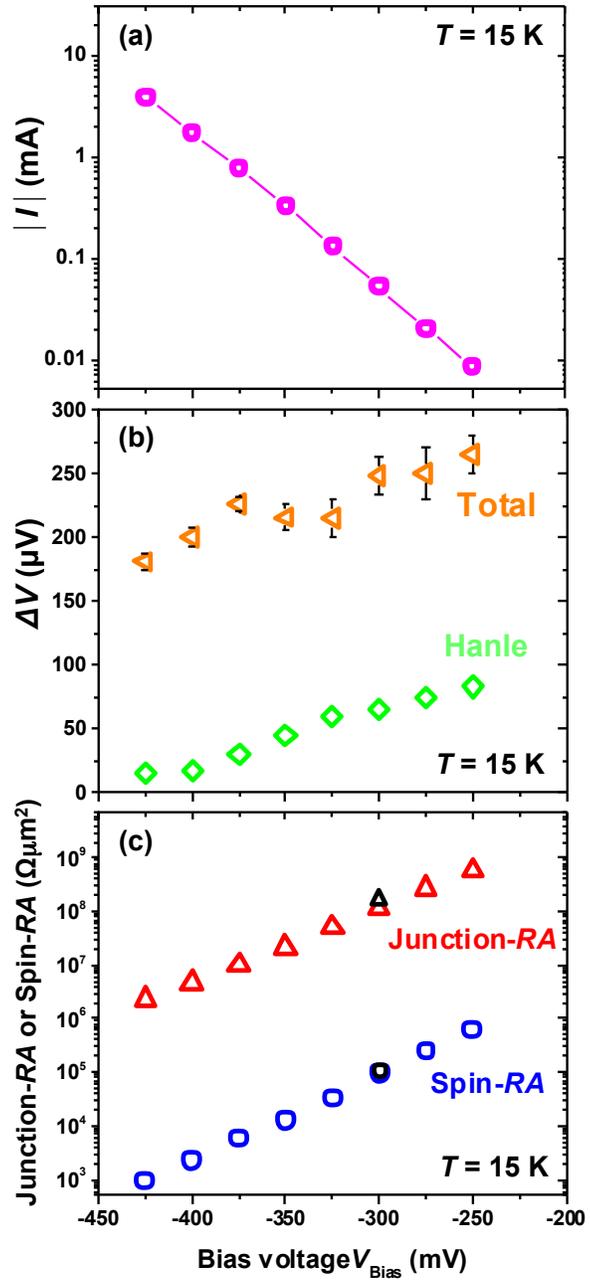

FIG. 6 (a), (b) and (c)